\documentclass[11pt,a4paper]{article}
\usepackage[nohead, nomarginpar, margin=1in, foot=.25in]{geometry}

\usepackage{graphicx}
\usepackage{color}

\usepackage{amsmath}
\usepackage{amsmath,bbm}

\usepackage{amssymb,bm}

\usepackage[hyphens]{url} %long url

\usepackage{siunitx} %for degree symbol
\usepackage{csquotes}
\usepackage{subcaption}
\usepackage{multirow}% http://ctan.org/pkg/multirow
\usepackage{tabularx} %auto table width
\usepackage{tabulary} %auto table width

\usepackage[T1]{fontenc}
\usepackage[utf8]{inputenc}
\usepackage{authblk}

\usepackage{soul}
\usepackage[LGRgreek]{mathastext}

\usepackage{url}

\providecommand{\keywords}[1]{\textit{Key words---} #1}
\providecommand{\pacs}[1]{\textit{JEL codes---} #1}

\usepackage{cite}
\date{}

\def\@listcomma@comma{\@ifnum{\@tempcnta>\tw@}{,}{}}

\begin{document}

\title{Impact of crop diversification on tribal farmer's income: A case study from Eastern ghats of India}
\author{Sadasiba Tripathy\footnote{tripathy.sadasiba@gmail.com}}% \footnote{Orcid id: 0000-0002-5275-8142}}
\author{Dr. Sandhyarani Das}
\affil{Berhampur University, Berhampur-760007, Odisha, India}

\maketitle

\begin{abstract}
In this investigation we analyze impact of diversification of agriculture on farmer's income, a study from primitive tribal groups from eastern ghats of India. We have taken crop diversification index to measure the extent and regression formalism to analyze the impact, of crop diversification. Descriptive statistics is employed to know the average income of the farmers, paired results of crop diversification index. We observed a positive impact on crop diversification in scheduled areas and investigated reasons where it did not work.
%A positive impact of crop diversification on the income (and hence standard of living) of the farmers is observed.
\end{abstract}

\noindent
\keywords{Diversification of agriculture, Tribal farming, Farm households, Agribusiness} \\
\pacs{Q12, Q13} \\

%---------------------------------------------------------------------Introduction----------------------------------------------------------------------------------
\section{Introduction}
\label{intro}

\subsection{Crop diversification and tribal farming}
Diversification of agriculture refers to shift of the production of a single crop to number of crops and from paddy to non-paddy crops, for example shift in the production pattern from the cereals to the high value crops \cite{Raoetal2008}. It reduces the risk involved in traditional agricultural practices to get better returns in modern agricultural practices \cite{Tripathy2019}. Agricultural diversification leads to increase in the standard of living of the people and increase in the demand of the high value crops \cite{Birthaletal2007}.
%The positive impact of Kissan credit card system and the process of financial inclusion in India by following the multiple regression analysis \cite{SumaiyaGandhimathi2015}.
%The diversification in agriculture will be helpful in raising the standard of living of the farmer in Tamilnadu \cite{VelavanBalaji2012}.
The factors like High yielding variety (HYV) seeds, irrigation, rainfall, size of the lands availability of market, are key factors that influence diversification of agriculture practices \cite{Acharjyaetal2011}. 

%Tribal people, often associated with deprivation from mainstream, like education, training or financial ability at par with modern civilization. As a policy of uplift them, government agencies take some steps to improvise their income and bring them to mainstream. For example, animal farming, fish cultivation etc \cite{}. 

%In Odisha, we can experience inclusive growth in last two decades and agriculture is the prime occupation of the people. The direct dependence on agriculture is around 48\% of the population of the state \cite{GoO2015}. As per the figure per month household income was INR 7,731 in 2016-17 from INR 4,976 in 2012-13\cite{OES201819}.

%\textcolor{red}{introduce tribal agriculture}
%\textcolor{red}{why they moved to traditional paddy to vegetable}
The primitive tribal groups (PTGs) \cite{ptgs} in Eastern Ghats of India used to cultivate millet  (Finger millet/Mandia, Foxtail millet/Kangu, Little millet/Suan) by Shift Cultivation (Podu cultivation) \cite{podu_0}. For rest of food requirements either they used hunting or depended on forest collections viz, edible roots (sweet patato), Bamboo shoots (Karadi), mahuwa flowers, tamarind etc. Podu cultivation caused heavy deforestation and added adverse effect to global warming \cite{podu}. So government agencies incorporated various steps \cite{podu_3} for suggesting them land based agriculture, which mostly was paddy (rice) until 2008-09 in the surveyed areas. %Few of them used to produce the vegetables in their backyard of their home for their home needs. 
Demonstration of few things convinced them for crop diversification viz.,benefit in less time, multiple time harvesting in single farming session, availability of market in local areas, possibility of higher price than traditional crops are few of them. The crop diversification started after that in small scale. Once the farmers get some short period benefits they started crop diversification to a larger extent. 

We choose Koraput (18.8135\si{\degree} N, 82.7123\si{\degree} E) district of state of Odisha, situated around 2nd major mountain peaks of Eastern ghat (1672 m) and at elevation of  870 m. Here one can see $\sim$ 84\% of tribal groups (52 tribes out of 62) seen in entire Odisha \cite{tribelist, tribelist_kpt}. Surveyed area chosen basing on two factors;  areas where at least 60\% of population dominated by tribal groups (Raniput/Village 1, Soralguda/Village2, Mundaguda/Village 3, Ekomba/Village 4 and Rupabeda/Village 5) and those areas have adopted crop diversification in recent years (13 such crops were considered). 

As primary data,  selection of village was done under purposive sampling according to the tribal concentration (ST). We had prepared structured interview schedule to collect the data of 80\% of the households. These households were chosen from each village by simple random sampling method. The time period chosen was 2008-09, 2013-14 and 2018-2019. We choose 2008-09 as base year. For the reason, only few crops were attempted before and too less people involved in diversification. And then we collected data for 5 years after that, for two terms; i.e. 2013 and in 2018.
Authors carefully surveyed those villages and collected primary data used for this analysis.

The secondary data was drawn from the statistical abstracts of Directorate of Economics \& Statistics (DES), Odisha \cite{dse_link} and Odisha Agriculture statistics 2013-14 \cite{agri_link}. For data analysis, we have taken crop diversification index \cite{Kamrajuetal2017} by taking into account as much as 13 crops of study areas.

%Objective of this study is to find the impact of crop diversification on tribal farmers income and to suggest measures to improve the process of crop diversification in these study areas.

%In this study, we choose tribal dominated regions of Odisha and such villages where the effect of diversification can be seen.

%The analysis is based on primary date collected by authors. 

\subsection{Profile of the Study Area}
%\label{intro_sec1}
We have considered such villages where tribal groups are more than 60\% and these villages should be small in size. Census data is represented in Tab. \ref{Tab:VillageCensus2011}.  The villages tribal population vary from 65\% to $\sim$ 100\% as per last census by Govt. of India \cite{CensusofIndia2011}.  Number of families in these villages vary from a minimum of 68 to 132 families, with all family members from  227 to  608 in maximum. Literacy goes as minimum of 16\% (for mostly tribal populated village) to 53.91\%; the upper limit even less than state average which is  72.87\% \cite{CensusofIndia2011}.

%ensus there are 103 families in Raniput with 495 population out of which females are 250 and 245 are males. It is a small village. The children population in Raniput village is 108 within the age group of 0-6. In the state Odisha the average sex ratio is 979 but in this village it is quite higher i.e. 1020.Coming to the literacy rate we can see it is at 38.76\% which is lower than the state's average i.e. 72.87\%. It can be explained in Tab. \ref{Tab:VillageCensus2011}.In the village 65.05\% of schedule tribe (ST) and 28.08\% of schedule caste (SC) people are there. So they are the dominated inhabitants of the land.

\begin{table}[h!]
\centering
\caption{Population as per 2011 census \cite{CensusofIndia2011}}
\begin{tabulary}{\textwidth}{L| L| L| L| L| L}  
   \hline
Particulars &Village 1 &Village 2 &Village 3 &Village 4 &Village 5 \\
   \hline
  No. of families &103 &122 &134 &100 &68 \\
     \hline
Total family members &495 &423 &608 &442 &227 \\
     \hline
Child (0-6 age) &108 &84 &130 &71 &41 \\
     \hline
     Schedule Tribe (in \%) &65.05 &64.77 &77.3 &80.77 &99.56 \\
     \hline
Schedule Caste (in \%) &20.08 &21.04 &0. &0.97 &0. \\
     \hline
Literacy (in \%) &38.76 &36.58 &35.36 &53.91 &16.13 \\
     \hline
\end{tabulary}
\label{Tab:VillageCensus2011}
\end{table}

To get information about the number of people being involved in crop diversification, we prepared a list of worker class. This can be found in Tab. \ref{Tab:WorkProfile2011}. Total workers who are involved in agricultural sector vary from a minimum of 144 to  317 maximum. Definition of main workers are those, who get the employment more than 6 months a year. And those who are engaged in works less than 6 months a year we call them marginal workers \cite{CensusofIndia2011}. Except one village, we see almost a 50-50 number for main worker and marginal worker.
%\subsection{Caste factor}
%\label{intro_sec2}
%In the village 65.05\% of schedule tribe (ST) and 28.08\% of schedule caste (SC) people are there. So they are the dominated inhabitants of the land.

%\subsection{Work profile of the Study Area}
%\label{intro_sec3}
%In Raniput Village 97 people describe them as main workers who get the employment more than 6 months a year and those who are engaged in works less than 6 months a year are called them marginal workers i.e. 118 numbers of people. The total workforce in Raniput village is 215.

\begin{table}[h!]
\centering
\caption{Work profile as per 2011 census \cite{CensusofIndia2011}}
\begin{tabular}{ c|c|c|c|c|c}  
   \hline
Particulars &Village 1 &Village 2 &Village 3 &Village 4 &Village 5 \\
   \hline
  Total Workers &215 &244 &317 &267 &144\\
     \hline
Main Worker (in \%) &45.12 &47.95 &47 &35.12 &56.94\\
     \hline
Marginal Worker (in \%) &54.88 &52.05 &53 &64.79 &43.06\\
     \hline
\end{tabular}
\label{Tab:WorkProfile2011}
\end{table}

\subsection{Area allocation for crop diversification}
\label{Areaallocationforcropdiversification}
One can get information about irrigated and non-irrigated areas  (in possession ) of a village from the Govt sources \cite{irrigtedarea_1, irrigtedarea_2}. However the actual land used for cultivation may vary for various reasons. Authors from their filed survey inquired about the actual land used for crop diversification and presented them (in \%) in Tab. \ref{Tab:AreaunderCrops}. From irrigated areas (apart from paddy harvesting session which is 60-90 days), one needs 45-60 days for vegetable harvesting to start. Depending upon vegetable, it goes 60-90 days harvesting time. 

%For individual crops, we listed amount of production per acre and presented them in Tab \ref{Tab:Productionofindividualcropsperacre}.

%Here we have chosen 3 time periods i.e. 2008-09, 2013-14 and 2018-19 \textcolor{red}{but as there were no such instances of crop diversification in the study area in 2008-09, so we have taken 2013-14 and 2018-19 to analyze the crop diversification}. 

%In Mundaguda village of Borigumma block, the farmers go for lowering the production of Sugarcane in 2013-14 as they faced a crop failure in 2011 i.e. the red 

\begin{table}[h!]
\centering
\caption{Area under Crops in \%}
\begin{tabulary}{\textwidth}{L| L| L| L| L| L| L}  
Year &Crop &Village 1 &Village 2 &Village 3 &Village 4 &Village 5\\
\hline
2008-09 &Paddy &100 &93.4 &83.61 &76.77 &100\\
    \cline{2-7}
~ &Spiny gourd &0 &0 &0 &0 &0\\
    \cline{2-7}
~ &Pointed gourd &0 &0 &0 &0 &0\\
    \cline{2-7}
    ~ &Coccinia &0 &0 &0 &0 &0\\
    \cline{2-7}
    ~ &Brinjal &0 &4.26 &3.42 &0 &0\\
    \cline{2-7}
    ~ &Cowpea &0 &0 &0 &0 &0\\
    \cline{2-7}
     ~ &Ladies finger &0 &2.3 &0 &0 &0\\
    \cline{2-7}
        ~ &Tamato &0 &0 &1.6 &1.62 &0\\
    \cline{2-7}
    ~ &Beans &0 &0 &0 &0 &0\\
    \cline{2-7}
        ~ &Cabbage &0 &0 &0 &0 &0\\
    \cline{2-7}
        ~ &Green peas &0 &0 &0 &0 &0\\
    \cline{2-7}
        ~ &Pumpkin &0 &0 &0 &2.17 &0\\
    \cline{2-7}
        ~ &Maize &0 &0 &0 &7.83 &0\\
    \cline{2-7}
    ~ &Sugarcane &0 &0 &11.37 &8.61 &0\\
    \cline{2-7}       
\hline
\hline
2013-14 &Paddy &48 &85.34 &78.89 &72.08 &91.28\\
    \cline{2-7}
 ~   &Spiny gourd &12.36 &0 &0 &0 &0\\
    \cline{2-7}
~ &Pointed gourd &9.84 &0 &0 &0 &0\\
    \cline{2-7}
    ~ &Coccinia &5.23 &0 &0 &0 &0\\
    \cline{2-7}
    ~ &Brinjal &5.76 &7.26 &4.92 &0 &8.72\\
    \cline{2-7}
    ~ &Cowpea &11.91 &0 &0 &0 &0\\
    \cline{2-7}
    ~ &Ladies finger &0 &3.27 &0 &0 &0\\
    \cline{2-7}
        ~ &Tamato &0 &0 &6.33 &4.35 &0\\
    \cline{2-7}
    ~ &Beans &0 &4.13 &0 &0 &0\\
    \cline{2-7}
        ~ &Cabbage &0 &0 &4.83 &0 &0\\
    \cline{2-7}
        ~ &Green peas &0 &0 &2.47 &0 &0\\
    \cline{2-7}
        ~ &Pumpkin &0 &0 &0 &2.82 &0\\
    \cline{2-7}
        ~ &Maize &0 &0 &0 &11.54 &0\\
    \cline{2-7}
    ~ &Sugarcane &6.9 &0 &2.56 &9.21 &0\\    
        \cline{2-7}  
 \hline
\hline
   2018-19 & Paddy &43  &73.77 &40.16 &63.4 &82.85\\
    \cline{2-7}
  ~ &Spiny gourd &18.6 &0 &0 &0 &0\\
    \cline{2-7}
~ &Pointed gourd &11.2 &0 &0 &0 &0\\
    \cline{2-7}
    ~ &Coccinia &6.21 &2.6 &0 &0 &5.17\\
    \cline{2-7}
    ~ &Brinjal &5.14 &8.42 &4.37 &0 &9.4\\
    \cline{2-7}
    ~ &Cowpea &9.62 &0 &0 &0 &2.58\\
    \cline{2-7}
      ~ &Ladies finger &0 &6.91 &0 &0 &0\\
    \cline{2-7}
        ~ &Tamato &0 &0 &6.42 &4.67 &0\\
    \cline{2-7}
    ~ &Beans &0 &8.3 &0 &0 &0\\
    \cline{2-7}
        ~ &Cabbage &0 &0 &9.43 &0 &0\\
    \cline{2-7}
        ~ &Green peas &0 &0 &7.25 &0 &0\\
    \cline{2-7}
        ~ &Pumpkin &0 &0 &0 &3.84 &0\\
    \cline{2-7}
        ~ &Maize &0 &0 &0 &14.52 &0\\
    \cline{2-7}
    ~ &Sugarcane &6.23 &0 &12.66 &13.57 &0\\     
\end{tabulary}
\label{Tab:AreaunderCrops}
\end{table}

\subsection{Harvesting and selling strategy in study area}
\label{HarvestingandSellingStrategy}
%We choose 2008-09 as base year. For the reason, only few crops were attempted before and too less people involved in diversification. And then we collected data for 5 years after that, for two terms; ie 2013 and in 2018.

%The most important part of this paper is to know the way the farmers harvest their diversified crops. 

We asked individual farmers and about their selling strategy and price they got for it. They use to pluck the vegetable crops (Spiny guard, Pointed guard Coccinia, Cowpea, Ladies finger, tomatto, beans) in every 3 days a week for 3 months and sell it in the nearby markets (either in wholesale or retail). For Brinjal and Green peas they use to pluck once in a week for 2 months. Coming to Sugarcane, maize, pumpkin, cabbage, these are one time harvesting products but you can harvest sugarcane once in a year for 3 years.  The farmers also sell Pumpkin plant's fruit, flower and leaf. In a good market the farmer's get a supporting price for the crops also depends on the spending habit of the people but it is not possible at some regions where the farmer's won't get a good market nearer to their villages. The crops like maize and sugarcane got a good market in recent years.  From their unit selling price and total amount of crop yield, we estimated the total income of the farmer. In Tab. \ref{Tab:Productionofindividualcropsperacre} we listed one can find crops yield amount per acre, for every single crop that we considered in this study.

\begin{table}[h!]
\centering
\caption{Diversified crop production (per acre, in quintals) }
\begin{tabulary}{\textwidth}{L| L| L| L| L| L| L}  
Year &Crop &Village 1 &Village 2 &Village 3 &Village 4 &Village 5\\
\hline
2008-09 &Paddy &NA &NA &NA &NA &NA\\
    \cline{2-7}
~ &Spiny gourd &0 &0 &0 &0 &0\\
    \cline{2-7}
~ &Pointed gourd &0 &0 &0 &0 &0\\
    \cline{2-7}
    ~ &Coccinia &0 &0 &0 &0 &0\\
    \cline{2-7}
    ~ &Brinjal &0 &120 &50 &0 &0\\
    \cline{2-7}
    ~ &Cowpea &0 &0 &0 &0 &0\\
    \cline{2-7}
     ~ &Ladies finger &0 &110 &0 &0 &0\\
    \cline{2-7}
        ~ &Tamato &0 &0 &80 &40 &0\\
    \cline{2-7}
    ~ &Beans &0 &0 &0 &0 &0\\
    \cline{2-7}
        ~ &Cabbage &0 &0 &0 &0 &0\\
    \cline{2-7}
        ~ &Green peas &0 &0 &0 &0 &0\\
    \cline{2-7}
        ~ &Pumpkin &0 &0 &0 &20 &0\\
    \cline{2-7}
        ~ &Maize &0 &0 &0 &9 &0\\
    \cline{2-7}
    ~ &Sugarcane &0 &0 &90 &40 &0\\
    \cline{2-7}       
\hline
\hline
2013-14 &Paddy &NA &NA &NA &NA &NA\\
    \cline{2-7}
 ~   &Spiny gourd &70 &0 &0 &0 &0\\
    \cline{2-7}
~ &Pointed gourd &60 &0 &0 &0 &0\\
    \cline{2-7}
    ~ &Coccinia &52 &0 &0 &0 &0\\
    \cline{2-7}
    ~ &Brinjal &70 &90 &60 &0 &51\\
    \cline{2-7}
    ~ &Cowpea &70 &0 &0 &0 &0\\
    \cline{2-7}
    ~ &Ladies finger &0 &110 &0 &0 &0\\
    \cline{2-7}
        ~ &Tamato &0 &0 &90 &45 &0\\
    \cline{2-7}
    ~ &Beans &0 &120 &0 &0 &0\\
    \cline{2-7}
        ~ &Cabbage &0 &0 &120 &0 &0\\
    \cline{2-7}
        ~ &Green peas &0 &0 &40 &0 &0\\
    \cline{2-7}
        ~ &Pumpkin &0 &0 &0 &22 &0\\
    \cline{2-7}
        ~ &Maize &0 &0 &0 &10 &0\\
    \cline{2-7}
    ~ &Sugarcane &120 &0 &120 &40 &0\\    
        \cline{2-7}  
 \hline
\hline
   2018-19 & Paddy &NA  &NA &NA &NA &NA\\
    \cline{2-7}
  ~ &Spiny gourd &92 &0 &0 &0 &0\\
    \cline{2-7}
~ &Pointed gourd &60 &0 &0 &0 &0\\
    \cline{2-7}
    ~ &Coccinia &52 &2 &0 &0 &70\\
    \cline{2-7}
    ~ &Brinjal &70 &90 &70 &0 &64\\
    \cline{2-7}
    ~ &Cowpea &85 &0 &0 &0 &60\\
    \cline{2-7}
      ~ &Ladies finger &0 &110 &0 &0 &0\\
    \cline{2-7}
        ~ &Tamato &0 &0 &55 &40 &0\\
    \cline{2-7}
    ~ &Beans &0 &120 &0 &0 &0\\
    \cline{2-7}
        ~ &Cabbage &0 &0 &80 &0 &0\\
    \cline{2-7}
        ~ &Green peas &0 &0 &40 &0 &0\\
    \cline{2-7}
        ~ &Pumpkin &0 &0 &0 &16 &0\\
    \cline{2-7}
        ~ &Maize &0 &0 &0 &9 &0\\
    \cline{2-7}
    ~ &Sugarcane &120 &0 &70 &50 &0\\     
\end{tabulary}
\label{Tab:Productionofindividualcropsperacre}
\end{table}

%The paper is organized as follows. We have introduced tribal farming system and the reasons for a change from their traditional farming in Introduction section. We will explain description of variables and model formalism in methodology section. This will followed by results and discussion. We will conclude the paper with policy suggestions in Sec IV.

%\subsection{Objectives}
%\label{intro_sec4}
%\begin{itemize}
%\item To study the impact of crop diversification on farmers income in the study area.
%\item To suggest measures to improve the process of crop diversification in the study area.
%\end{itemize}

%\section{Analysis Technique}
\section{Methodology}

\subsection{Description of variables}
\label{variables}
We measured the crop diversification index through simple linear regression (SLR) model.  It examines impact of crop diversification on farmer's income in the study area.

Crop diversification Index (CDI) as given by Gibbs and Martin's technique \cite{GMtechnique1962}:
\begin{equation}
CDI = \frac{\sum X^2}{(\sum X)^2}
\end{equation}

Where X = area under individual crops (in \%).\\
The value of CDI lies between zero and one; zero indicates no diversification and unity states cent percent diversification. We have taken 2 variables: Crop Diversification, Farmer's income %\textcolor{red}{and standard of living}.

Crop Diversification:\\
Crop diversification is considered as one of important parameters of growth of an economy \cite{Brithaletal2006}. In the study area the farmer's started crop diversification from paddy to vegetables in the recent years. Here, we have taken the area diversified for crop from a traditional practice of paddy to vegetables. It is used as the independent variable.

Farmer's income:\\
The income plays a very important role for the further inducement for investment and to take risks \cite{Nayak2015}. Here we have taken income as the dependent variable. %\textcolor{red}{It can be expected that there is a positive impact of increased in income on crop diversification}.

\subsection{The Model}
\label{model}
Here we have taken crop diversification as the independent variable and farmer's income as dependent variable. Here, we would analyze change in the crop diversification leads to the change in farmer's income.\\
Y = $b_0+b_1X_1+\epsilon$ \cite{linearregessioneqn}\\
Where, Y = income (dependent variable),\\
$b_0$ = estimate of the regression intercept,\\
$b_1$ = estimate of the regression slope,\\ 
$X_1$ = crop diversification (Independent variable),\\
and $\epsilon$ is error term.

%\subsection{Crop production fluctuation in different regions in different years}
%In Mundaguda village of Borigumma block, the farmers go for lowering the production of Sugarcane in 2013-14 as they faced a crop failure in 2011 i.e. the red root problem. The farmers faced a huge loss and stopped producing sugarcane except to meet the need of the local markets for molasses.

%%---------------------------------------------------------------------Results----------------------------------------------------------------------------------
\section{Results and discussion}
\label{Results_and_discussion}

\subsection{Linear regression analysis}
In Tab. \ref{Tab:Regressionanalysis} we have calculated regression variables for the year 2013-14 and 2018-19 along with base year 2008-09. % As we have mentioned earlier, no crop diversification seen in the years of 2008-09, so we did not present here.

Standard Deviation (SD) of the mean value (by direct method) may be calculated as:
 \begin{equation}
  \sigma = \sqrt{\frac{\sum x^2}{N}}
 \end{equation}
Where x = difference of value of X from it's mean and N is number of data points.

\begin{table}[h!]
\centering
\caption{Descriptive Statistics }
%\begin{tabular}{ c|c|c|c|c|c|c}  
\begin{tabulary}{\textwidth}{L| L| L| L| L| L| L}  
Variables  &Year & Village 1 &Village 2 &Village 3 &Village 4 &Village 5\\
\hline
CDI(X) &2008-09  &0 &0.544635 &0.534312 &0.348866 &0\\
    \cline{2-7}
~ &2013-14    & 0.184757 & 0.374367 &0.224981 &0.314127 & 1.\\ 
    \cline{2-7}
  ~  &2018-19    & 0.205522  & 0.2824 &0.224834 &0.322143 & 0.413927\\ 
    \cline{2-7}
~ & Mean CDI &0.130093 &0.400467 &0.328042 &0.328379 &0.471309\\
    \cline{2-7}
~ & SD of CDI &0.0923794 &0.108636 &0.145855 &0.0148518 &0.41026\\
\hline
\hline
 Income (Y) (in crores) &2008-09 &0. &0.14 &0.27 &0.32 &0  \\
 \cline{2-7}
 ~  &2013-14 &0.97 &0.36 &0.64 &0.52 &0.0247 \\ 
       \cline{2-7}
~&2018-19 &1.82 &0.71 &1.03 &0.86 &0.063 \\ 
       \cline{2-7}
~ & Mean Income &0.93 &0.403333 &0.646667 &0.566667 &0.0292333\\
    \cline{2-7}
~ & SD of Income &0.74355 &0.23471 &0.310304 &0.22291 &0.0259186 \\
%   \end{tabular}
\end{tabulary}
\label{Tab:Regressionanalysis}
\end{table}

A simple linear regression (SLR) may be written as:
\begin{align}
Y &= b_0+b_1X \\
where \
b_0 &= \bar{Y} - b_1 \bar{X}, \\
and \
b_1 &= \frac{N\sum XY - \sum X \sum Y}{N\sum X^2 - (\sum X)^2}
\end{align}

Application of SLR to the descriptive statistics can be found in Tab. \ref{Tab:RegressionEquation}. Here we have calculated SLR to all five village data. %Interestingly, all village data shows consistently increase of income variable of tribal farmers.
\begin{table}[h!]
\centering
\caption{Regression Equation}
\begin{tabular}{ c|c|c|c}  
   \hline
Village Name &$b_1$ &$b_0$ &$Y = b_0+b_1X$ \\
\hline
Village 1 &7.43324 &-0.0370122 & $Y= -0.0370122+7.43324 \times X$ \\
\hline
Village 2 &-2.06297 &1.22949 &$Y= 1.22949+ -2.06297  \times X$ \\
\hline
Village 3 &-1.82654 &1.24585 &$Y= 1.24585+ -1.82654 \times X$  \\
\hline
Village 4 &-9.3954 &3.65192 &$Y= 3.65192+ -9.3954 \times X$  \\
\hline
Village 5 &0.0187025 &0.0204187 &$Y= 0.0204187+ 0.0187025 \times X$  \\
\hline
\end{tabular}
\label{Tab:RegressionEquation}
\end{table}

\begin{figure}
\centering
\includegraphics[scale=0.5]{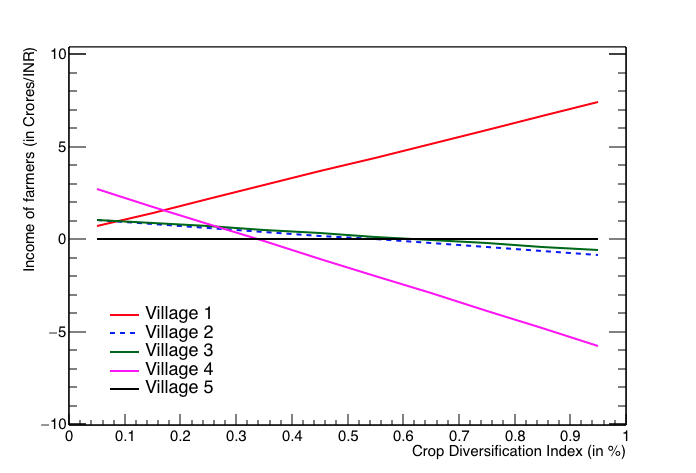}
\caption{(Color online) Income of farmers as a function of CDI, calculated from Linear regression.}
\label{fig:CDIVsIncome} 
\end{figure}

In Fig. \ref{fig:CDIVsIncome} we have plotted regression equation with CDI as mentioned in Tab. \ref{Tab:RegressionEquation}. 
%This shows a sharp linear increase of income of farmers, when CDI goes increasing. 
The figure explains the change of income of farmers by crop diversification in the study area. In Village 1 and Village 5, we can see there is an increasing trend (Village 5 change is not visible in the figure). The straight line speaks that there has been an increase, on the income of the farmers, in Village 1. The factors that supported for this positive impact can be a good nearby market i.e. Jeypore market, which is < 6km (Boipariguda market for Village 5, which is < 10kms)where the demand for the green vegetables is quite high supported by a better road communication.
 In the villages like Village 2 and Village 3, though there are instances of crop diversification but it hasn't made a remarkable impact on the income of the farmer's. There are two reasons responsible for this. In the first place they don't have a good market and lack of a storage facility. So this made difference. In Village 4, we can experience a much decreasing trend because the village lacks low communication facility and lack of supporting price.

%\begin{equation}
%Y = b_0+b_1X
%\label{eq:Eq1}
%\end{equation}
%where
%\begin{equation}
%b_0 = \bar{Y} - b_1 \bar{X}
%\label{eq:Eq2}
%\end{equation}

% \begin{align}
%b_1 &= \frac{N\sum XY - \sum X \sum Y}{N\sum X^2 - (\sum X)^2} \\
%&=\frac{10 \times 0.5386 - 0.38 \times 2.79}{10 \times 0.0724 - {0.38}^2} \nonumber \\
%&=7.59 \nonumber \\
%\end{align}
%So Eq. \ref{eq:Eq2} modified as following \\
% \begin{align}
%b_0 &= 1.39 - 7.59 \times 0.19 \\
%&=-0.05\\
% \end{align}
%Hence Eq. \ref{eq:Eq1} modified as following \\
% \begin{align}
% Y = -0.05 + 7.59 \times X
%  \end{align}
  
% \begin{table}[h!]
%\centering
%\caption{Impact of Crop diversification on Income}
%\begin{tabular}{ c|c|c}  
%   \hline
%   Year & 2013-14 & 2018-19 \\
%   \hline
%Crop diversification &0.18 &0.2\\
%   \hline
%Income (in crore) &0.97 &1.82\\
%   \hline
%\end{tabular}
%\label{Tab:ImpactofCropdiversificationonIncome}
%\end{table}

%%---------------------------------------------------------------------Summary----------------------------------------------------------------------------------
\section{Conclusion and policy recommendations} 
\label{Summary}
We have investigated effect on tribal farmers income from crop diversification. Farmers data taken from such places where tribal population is dominated in South Odisha. From the area used for farming we observe that, there is increasing allocation of land for crop diversification from traditional paddy (rice or ragi) to vegetables (13 such we have analyzed). For the year 2008-09, where almost zero land used for vegetable crop, almost 50\% of land used in the years 2013-14 and 2018-19. Hence CDI increases to a factor 11\%. In case of some vegetables the extent of the crop diversification is very high but on an average it is very slow. Regression analysis suggests that, the crop diversification has a positive impact on the farmer's income. This findings is consistent in all five areas where the data samples were taken.

Limitations of this study, is that we did not consider crop diversification or production, for house kitchen purposes. This factor although do not represent the amount of yield to be considered as crop diversification. Irrigation facility and availability of fertilizer \& pesticides can be added constraint of the linear regression formalism we have considered. However most of tribal people in study area adopts organic farming and the cropped areas are rainfed only. So we safely ignored this constraint as well. Although climatic condition and soil type can be change the numbers represented here, but that's beyond of someone's control.

% The result of the paper can be verified at macro level for the wider scope.

%From our field study, we propose few things for the growth of the agricultural practice in the study area. 
On policy suggestions, authors found and hence suggest that few things for further improvements. 
One major concern is to provide a supportive price and mini store houses. Since vegetables get rotten in 3-4 days, so famers forced to sell them in lesser price without having a storage facility. HYV seeds as per the climatic condition will be added benefit for them.  Also learning programs should be provided to the farmers in their villages through Digital Green System where short videos of success stories should be screened.

%--------------------------------------------------------------------- Reference----------------------------------------------------------------------------------

\end{document}